\documentclass[prd,superscriptaddress,amsfonts,amssymb,amsmath,showpacs,twocolumn,nofootinbib]{revtex4-2}
\usepackage{bm}
\usepackage{amsfonts}
\usepackage{latexsym}
\usepackage{graphicx}
\usepackage{amsmath}
\usepackage{palatino}
\usepackage{xcolor} 
\usepackage{mathpazo}
\usepackage{tensor}
\usepackage{textcomp}
\usepackage{orcidlink}
\usepackage{float}
\usepackage{booktabs}
\usepackage{dcolumn}
\usepackage{booktabs}
\usepackage{multirow}
\usepackage{hyperref}
\hypersetup{colorlinks,citecolor=blue}
\usepackage{amsmath}
\usepackage{xcolor}
\usepackage{orcidlink}
\usepackage{epsfig}
\usepackage{caption}
\usepackage{subcaption}
\usepackage{commath}
\hypersetup{colorlinks,citecolor=blue}
\hypersetup{colorlinks=true,linkcolor=magenta,filecolor=magenta,    urlcolor=blue}

\def\be{\begin{equation}}
\def\ee{\end{equation}}
\def\bea{\begin{eqnarray}}
\def\eea{\end{eqnarray}}

\begin{document}

\title {\bf {Joint constraints on $R_h=ct$ cosmology from DESI DR2 BAO, CC, and SN\textit{Ia} Pantheon$^+$ sample}}

\author{Amritansh Mehrotra \orcidlink{0009-0006-0601-6603}}
\email{mhtra.amritansh@gmail.com}
\affiliation{Department of Physics, University Institute of Sciences, Chandigarh University, Mohali 140413, India}
\affiliation{Pacif Institute of Cosmology and Selfology (PICS), Sagara, Sambalpur 768224, Odisha, India}

\author{S. K. J. Pacif \orcidlink{0000-0003-0951-414X}}
\email{shibesh.math@gmail.com}
\affiliation{Pacif Institute of Cosmology and Selfology (PICS), Sagara, Sambalpur 768224, Odisha, India}
\affiliation{Research Center of Astrophysics and Cosmology, Khazar University, Baku, 41 Mehseti Street, AZ1096, Azerbaijan}

\author{A. F. Santos \orcidlink{0000-0002-2505-5273}}
\email{alesandroferreira@fisica.ufmt.br}
\affiliation{Programa de P\'{o}s-Gradua\c{c}\~{a}o em F\'{\i}sica, Instituto de F\'{\i}sica,\\ 
Universidade Federal de Mato Grosso, Cuiab\'{a}, Brasil}

\begin{abstract}
    We carry out a comparative analysis of the standard $\Lambda$CDM cosmological model and the alternative $R_h=ct$ framework using recent observational data from cosmic chronometers (CC), Type Ia supernova, and baryon acoustic oscillations. The study evaluates the ability of each model to reproduce the observed expansion history of the Universe through a joint statistical assessment based on $\chi^2$ statistics, Akaike Information Criterion $(AIC)$, Bayesian Information Criterion $(BIC)$, and Bayes factor. While both models yield acceptable fits, $\Lambda$CDM consistently attains lower information-criterion values and higher likelihood, indicating a superior overall performance. An examination of the redshift evolution of the Hubble parameter $H(z)$ and the deceleration parameter $q(z)$ shows that $\Lambda$CDM naturally captures the transition from early-time deceleration to late-time acceleration, where as $R_h=ct$ predicts a strictly linear expansion. We also estimate the age of the Universe within both models, obtaining $t_0^{\Lambda CDM}= 13.676_{-0.81}^{+0.92}$Gyr and $t_0^{R_h=ct}= 16.035_{-0.98}^{+1.09}$Gyr. The posterior-derived age in the $\Lambda$CDM framework is broadly consistent with the Planck 2018 CMB result. This agreement is interpreted as a validation of the analysis pipeline and the reliability of the DESI DR2, CC, and supernova constraints, rather than as a new result for $\Lambda$CDM, and serves as a benchmark for assessing the viability of the $R_h=ct$ model. Recent JWST observations of unexpectedly mature high-redshift galaxies have renewed discussion regarding the timeline of early structure formation; although these results remain under active investigation, they underscore that fully resolving cosmic evolution may require refinements beyond the concordance paradigm. 
\end{abstract}
\maketitle

\section{Introduction}\label{sec_1}
The discovery of the late-time accelerated expansion of the Universe marks one of the most significant milestones in modern cosmology. Observations of high-redshift Type Ia supernovae (SNe Ia) provided compelling evidence that the expansion rate of the Universe is increasing rather than slowing down~\cite{Riess1998,Perlmutter1999}. Subsequent refinements in cosmological measurements~\cite{Jha2007,Guy2007,Burns2010} have further strengthened this conclusion, placing cosmic acceleration at the core of contemporary cosmological research. Complementary evidence from the cosmic microwave background radiation (CMBR) and large-scale structure (LSS) surveys~\cite{Mould2000,Spergel2003,Jarosik2011,Wood2007} further supports this picture.

The observational evidence suggests the existence of a form of energy with negative pressure, termed dark energy (DE), that drives the accelerated expansion~\cite{Turner2007}. Within the standard cosmological framework, this leads to a spatially flat Universe dominated by two unseen components, dark matter (DM) and dark energy (DE), with present-day energy density parameters of $\Omega_{m_0} \approx 0.31$ and $\Omega_{\Lambda_0} \approx 0.69$, respectively~\cite{collaboration2020planck}. Yet, its physical identity is unknown, and numerous theoretical proposals have emerged attempting to explain it. These include scalar field models~\cite{Ratra1988, Cladwell1998}, fluid-based scenarios~\cite{Kamenshchik2001,Bento2002}, and modified gravity theories~\cite{Carroll2004,Nojiri2011}, each aiming to reproduce the observed evolution of the Universe through distinct underlying mechanisms. Observational datasets such as SNe Ia, baryon acoustic oscillations (BAO), CC, and strong lensing systems (SLS)~\cite{Burns2010} have provided increasingly precise constraints on these models, but none have yet resolved the fundamental mystery.

The simplest and most widely accepted framework for explaining the accelerated expansion is the $\Lambda$CDM model, which incorporates a cosmological constant ($\Lambda$) with an equation of state $\omega = -1$ alongside cold dark matter with $\omega = 0$. This model successfully describes a broad range of cosmological observations, from early-Universe fluctuations to the present-day expansion rate. However, despite its empirical success, the $\Lambda$CDM model faces several persistent theoretical and observational challenges. On small scales, it struggles with the missing satellite and core–cusp problems~\cite{Bernardis2000,Hanany2000,Netterfield1997}, while on a fundamental level it suffers from the so-called cosmological constant problem, referring to the $10^{120}$-order discrepancy between the theoretically predicted and observed values of vacuum energy density~\cite{Mould2000,Spergel2003,Jarosik2011}. Moreover, the degeneracy issue between dark matter and dark energy components complicates the task of isolating their individual contributions to the cosmic energy budget. Furthermore, recent observational results have begun to hint at possible deviations from a simple cosmological constant description of dark energy. In particular, analyses of the latest baryon acoustic oscillation measurements from the Dark Energy Spectroscopic Instrument (DESI) Data Release 2 (DR2) have suggested that the dark energy equation of state may evolve with time, motivating further investigation using updated datasets~\cite{desicollaboration2025desidr2resultsii}.

To address these long-standing tensions, numerous alternatives to $\Lambda$CDM have been proposed, seeking to explain the observed acceleration without invoking a finely tuned cosmological constant. These range from dynamical dark energy models such as quintessence~\cite{Zlatev1999,Brax2000,Barreiro2000}, k-essence~\cite{k1999,Picon2000,Chiba2000}, and phantom fields~\cite{Caldwell2002}, to unified dark sector models such as the Chaplygin gas~\cite{Gorini2003}. Modified gravity frameworks, including $f(R)$ theories and braneworld scenarios~\cite{Capozziello2002,Dvali2000}, offer an alternative viewpoint in which cosmic acceleration arises from deviations in the gravitational sector at large scales. Several of these models have been tested against observational datasets such as CC, SNe Ia, BAO, and CMB measurements, with some providing viable phenomenological descriptions of cosmic acceleration, although many remain observationally degenerate with or mildly disfavored relative to the standard $\Lambda$CDM framework~\cite{Copeland2006,mehrotra2025}. Despite their conceptual diversity, these models collectively underscore the growing recognition that the standard $\Lambda$CDM framework, while observationally successful, may not represent the final description of cosmic evolution. Recent years have also witnessed increasing discussion of possible tensions within the standard cosmological framework. In particular, the persistent discrepancy between early-Universe measurements of the Hubble constant derived from CMB observations and late-time determinations based on local distance ladders, commonly referred to as the Hubble tension, has motivated renewed scrutiny of the assumptions underlying the $\Lambda$CDM model. In addition, the fundamental physical nature of dark matter and dark energy remains unknown. These open questions have encouraged continued exploration of alternative cosmological scenarios that may provide different perspectives on the expansion history of the Universe.

Beyond models that modify the dark-energy sector or gravitational physics, a different line of thought emerges from \textbf{coasting cosmologies}, in which the Universe expands linearly with cosmic time. The earliest and most well-known example is the Milne model~\cite{Milne1935}, developed by Arthur Milne, which is formulated as an empty, open Universe whose expansion is governed entirely by initial kinematics rather than gravitational effects. In this model the absence of matter, radiation, and a cosmological constant leads to a linear expansion that arises trivially from the lack of gravitational interactions. Although the Milne cosmology cannot describe the observed matter content and structure of the real Universe, it introduced the important notion that a constant expansion rate may arise naturally without invoking dark energy.

Modern coasting cosmologies differ conceptually from the original Milne framework. In particular, the $R_h=ct$ model is formulated within the framework of general relativity and assumes that the gravitational horizon satisfies the relation $R_h=ct$, which corresponds to the condition $\rho + 3p = 0$ for the total cosmic fluid. This requirement leads to a linear expansion history while still allowing the presence of matter and radiation components. In this sense, the $R_h=ct$ cosmology is not an empty Milne Universe but rather a dynamically constrained cosmological solution.

This idea has resurfaced in more recent discussions questioning whether the conventional division between geometry and energy-momentum is fundamentally necessary. In particular, critiques by Vishwakarma~\cite{Vishwakarma2013,Vishwakarma2016} have argued that a fully geometric approach to gravitation might eliminate the need for hypothetical dark sectors, thereby motivating further investigation into cosmological models exhibiting linear expansion behavior. In this context, coasting frameworks such as the $R_h=ct$ scenario offer an intriguing alternative worth confronting with contemporary observational data. Proponents of the $R_h=ct$ framework argue that a linear expansion history may arise from the global condition $\rho + 3p = 0$ for the total cosmic fluid, which ensures that the gravitational horizon satisfies $R_h = ct$. Although the standard cosmological picture describes successive radiation-, matter-, and dark-energy-dominated epochs that lead to a non-linear expansion history, it has been suggested that the $R_h=ct$ framework may provide an alternative description in which the overall expansion remains linear while still reproducing several key cosmological observables. This possibility motivates continued investigation of such models and their confrontation with modern observational datasets. This is further discussed in the following section.

This paper is organized in the following manner. Section~\ref{sec_1} gives a brief introduction. Section~\ref{sec_2} provides an overview for $R_h=ct$ cosmology. Section~\ref{sec_3} presents the datasets employed in this study, along with the statistical methods used to compare the two cosmological models. This section also includes a discussion of the Hubble parameter, the Deceleration Parameter (DP), the cosmic age, and consistency with the DESI DR2 observables which are further analyzed to strengthen the conclusions of the paper. Section~\ref{sec_4} summaries our findings after comparing the two models. Section~\ref{sec_5} provides our concluding remarks.

\section{The $R_h=ct$ Cosmology}\label{sec_2}
An intriguing alternative to the standard $\Lambda$CDM model is the $R_h = ct$ cosmology, first proposed and developed by Melia and collaborators~\cite{Melia2012,melia2012cosmic}. This model is built upon the condition that the gravitational horizon, defined as $R_h(t) = c / H(t)$, evolves in such a way that $R_h(t) = ct$ at all cosmic times. In general cosmological models the Hubble radius does not necessarily coincide with the particle horizon, which represents the maximum distance that light could have traveled since the Big Bang. However, within the $R_h = ct$ framework the linear expansion implies that the Hubble radius grows proportionally with cosmic time, leading to the relation $R_h(t) = ct$. This condition leads to a linear evolution of the scale factor, expressed as $a(t) \propto t$, implying a constant expansion rate.

From this relation, the Hubble parameter takes the form
\begin{equation}
H(z)=H_0(1+z),  \label{eq_1}
\end{equation}
where $H_0$ is the present-day Hubble constant and $z$ is the redshift. This expression corresponds to a cosmological evolution that is dynamically simpler than that of $\Lambda$CDM because it does not require distinct epochs of radiation, matter, and dark energy domination. Instead, the entire expansion history is governed by the condition that the total equation of state satisfies $\rho + 3p = 0$, which ensures that the active gravitational mass density of the Universe vanishes at all times.

{The $R_h = ct$ model has been tested against a variety of cosmological observations. Melia and collaborators have reported that this model can provide fits comparable to those of $\Lambda$CDM when confronted with diverse datasets, including CC~\cite{Melia2013,Melia2018}, quasar core angular size measurements~\cite{Wan2019}, X-ray and UV fluxes from quasars~\cite{Melia2019}, galaxy cluster gas mass fractions~\cite{Melia2016}, Type Ia supernovae~\cite{Melia2018b}, strong gravitational lensing statistics~\cite{Leaf2018,Melia2023}, fast radio bursts (FRBs)~\cite{wei2023investigating}, and recent JWST observations of high-redshift quasars and Einstein rings~\cite{melia2025lens,Melia2024}. In several of these studies the reported preference for the $R_h = ct$ framework is driven in part by fits to CC measurements of $H(z)$. However, CC determinations depend on stellar population synthesis modelling and may therefore be affected by systematic uncertainties~\cite{moresco2020setting}. The robustness of the inferred preference can also depend on whether CC data are used alone or in combination with other probes such as supernovae or BAO. Furthermore, in supernova-based comparisons the outcome may depend on the statistical assumptions adopted for the competing models. Allowing additional freedom in $\Lambda$CDM, for example by treating the curvature parameter $\Omega_{K}$ as a free parameter, increases the number of degrees of freedom and leads to stronger penalties in model selection criteria such as AIC and BIC, which explicitly depend on the number of free parameters \cite{akaike1974,bayesian1978}.

Several independent studies~\cite{Bilicki2012,Lewis2016,Haridasu2017,Lin2018,Hu2018,Tu2019,Fujii2020,Singirikonda2020} have also reported that the $R_h = ct$ framework is statistically disfavored when tested against combined observational probes or rigorous Bayesian model selection criteria. These differing conclusions often arise from methodological differences, including the choice and combination of datasets, the treatment of nuisance parameters, the adopted priors, and the statistical criteria used for model comparison. Although the simplicity of the model remains conceptually appealing, questions regarding its theoretical foundation and consistency with structure formation continue to motivate further investigation.
Raffai et al. have also tested coasting cosmological models, including the $R_h=ct$ model, using independent cosmological datasets~\cite{Raffai:2023ymr,Raffai:2024xnn,kodmon2025testing}.} In the next section we describe the datasets employed in the analysis and outline the statistical framework used to compare the two cosmological models. We also examine the redshift evolution of the Hubble parameter ($H(z)$) and DP ($q(z)$) together with the discussion of cosmic age and consistency of the models with the DESI DR2 observables.

\section{Datasets and Statistical Methods}\label{sec_3}
The parameters of models under study are constrained using the latest Baryon Acoustic Oscillation data from the DESI DR2 along with the latest measurements from Pantheon$^+$ and CC. For each dataset, the construction and sampling of the likelihood function is done using the nested sampling algorithm implemented in the \texttt{dynesty} package\footnote{\protect\url{https://dynesty.readthedocs.io/en/v2.1.5/index.html}}\cite{speagle2020dynesty}. This approach provides the combined benefits of efficiently exploring the posterior distribution and calculating Bayesian evidence at the same time, making it highly effective for both parameter estimation and model comparison. The analysis is performed by adopting uniform priors over physically motivated parameter ranges. For the $\Lambda$CDM model, the Hubble constant is varied over $H_0 \in [55,100]$, the matter density parameter over $\Omega_m \in [0,1]$, and the sound horizon at the drag epoch over $r_d \in [100,300]$. For the $R_h = ct$ model, uniform priors are assumed for $H_0 \in [55,100]$ and $r_d \in [100,300]$.
For each model, we employ 300 live points to sample this space. The resulting samples are then used to determine credible intervals and to visualize the constraints on the model parameters. To visualize the results, we make use of the \texttt{GetDist} library\footnote{\protect\url{https://getdist.readthedocs.io/en/latest/}} \cite{lewis2019getdist}, which produces detailed corner plots illustrating both the marginal posterior distributions and the correlations between model parameters. The full posterior distribution for each model is obtained by constructing dedicated likelihood functions for the corresponding datasets. The parameters of each model are constrained through a joint likelihood analysis incorporating all the datasets mentioned above. The overall likelihood function is expressed as
\begin{equation}
    \mathcal{L}_{Total}=\mathcal{L}_{CC}\times\mathcal{L}_{SNe~Ia}\times\mathcal{L}_{BAO}.
\end{equation}
The total chi-squared function, given as $\chi^2=-2\ln\mathcal{L}_{Total}$, is minimized to obtain the constraints on each model parameter. 
The datasets mentioned above are described as follows.
\begin{itemize}
    \item \textbf{Cosmic Chronometers (CC): }In this work, the first dataset used in this study is the CC compilation to constrain the parameters of the model under study. Specifically, we adopt 15 Hubble parameter measurements selected from a total of 31 available data points, covering the redshift range $0.1791 \leq z \leq 1.965$~\cite{moresco2012new,moresco2015raising,moresco20166}. These measurements are based on the differential age method, which estimates the Hubble parameter directly from the age evolution of massive, passively evolving early-type galaxies. Since these galaxies have not experienced significant star formation since their early formation epochs, their age difference at close redshifts provides a reliable and model-independent estimate~\cite{jimenez2002constraining} of the expansion rate through the following relation
    \begin{equation}
        H(z)= -\frac{1}{(1+z)}\frac{dz}{dt}.
    \end{equation}
    For our analysis, we employ the CC likelihood function developed by Moresco, available in his public GitLab repository\footnote{\protect\url{https://gitlab.com/mmoresco/CCcovarience}}. This likelihood incorporates the full covariance matrix, properly accounting for both statistical and systematic uncertainties~\cite{moresco2018setting,moresco2020setting}. By including these correlations, the method provides more robust and unbiased constraints on cosmological parameters. Recently, new CC measurements have also been derived using DESI DR2 observations, further highlighting the relation between BAO measurements from DESI and CC constraints in probing the expansion history of the Universe~\cite{loubserDESI}
    \item \textbf{Type Ia Supernova: }The second dataset used in this study is the Pantheon$^+$ compilation, which consists of light-curve measurements from 1701 Type Ia supernovae (SNe Ia) that span the redshift range $0.001 \leq z \leq 2.261$~\cite{scolnic2022pantheon+}. In this work we use the Pantheon$^+$ sample alone, without the SH0ES distance-ladder calibration. Since low redshift observations are subjected to significant systematic uncertainties caused by peculiar velocity effects, for this study we exclude the observations with $z < 0.01$. The likelihood used in our analysis is obtained from the public Pantheon$^+$ repository\footnote{\url{https://github.com/PantheonPlusSH0ES/DataRelease}} that includes the full covariance matrix expressed as $C_{tot}= C_{sys}+C_{stat}$. This formulation consistently incorporates both systematic and statistical uncertainties, ensuring a rigorous and reliable treatment of the supernova data~\cite{conley2011supernova}.
    \item \textbf{Baryon Acoustic Oscillations: }The third and final dataset that we incorporated in this study are 13 recent Baryon Acoustic Oscillation (BAO) measurements from the DESI DR2~\cite{karim2025desi}, covering the redshift range $0.295 \leq z \leq 2.330$. These measurements are derived from multiple galaxy and quasar samples\footnote{\url{https://github.com/CobayaSampler/bao_data}} and probe the large-scale clustering of matter in the Universe. BAO observations constrain cosmological models through several distance measures that encode both radial and transverse information of the large-scale structure. In particular, the DESI measurements are expressed in terms of three key distance indicators: the Hubble distance $D_H(z)$, the comoving angular diameter distance $D_M(z)$, and the volume-averaged distance $D_V(z)$.
    The Hubble distance is defined as
    \begin{equation}
    D_H(z)=\frac{c}{H(z)}, \label{Dh_eq}
    \end{equation}
    where $H(z)$ is the Hubble parameter and $c$ denotes the speed of light. The comoving angular diameter distance is given by
    \begin{equation}
    D_M(z)=c\int_0^{z}\frac{dz'}{H(z')}. \label{Dm_eq}
    \end{equation}
    The volume-averaged BAO distance combines radial and transverse information and is defined as~\cite{eisenstein2005bao}
    \begin{equation}
    D_V(z)=\left[z\,D_M^2(z)\,D_H(z)\right]^{1/3}.
    \end{equation}
    For comparison with theoretical predictions, we use the corresponding dimensionless ratios $D_H(z)/r_d$, $D_M(z)/r_d$, and $D_V(z)/r_d$, where $r_d$ denotes the sound horizon at the drag epoch (approximately $z\sim1060)$. For the case of $\Lambda$CDM model, this value is $r_d=147.09 \pm 0.26~Mpc$~\cite{collaboration2020planck}. However, to maintain model independence and accommodate possible deviation from standard recombination physics, we treat $r_d$ as a free parameter in our analysis~\cite{pogosian2020recombination,pogosian2024consistency,jedamzik2021reducing,lin2021early,vagnozzi2023seven}. Treating $r_d$ as a free parameter introduces a well-known degeneracy with the Hubble constant $H_0$ in BAO analyses, since BAO measurements primarily constrain the combination $H_0 r_d$ rather than $H_0$ or $r_d$ individually. Consequently, BAO data alone cannot tightly determine the value of $H_0$ when $r_d$ is allowed to vary. In the present analysis, this degeneracy is partially broken by the inclusion of CC measurements, which directly probe the expansion rate $H(z)$ and therefore provide additional constraints on $H_0${\color{red}~\cite{moresco20166,moresco2018setting,verde2019}}. The joint analysis of CC, Pantheon+, and DESI DR2 datasets therefore allows us to obtain meaningful bounds on $H_0$ while maintaining model independence in the treatment of the sound horizon.
    
\end{itemize}

To evaluate the relative performance of the model with respect to the standard $\Lambda$CDM
model, we compute the logarithm of the Bayes factor, defined as
\begin{equation}
    \ln B_{\Lambda i}= \ln \left( \frac{p(d \mid M_i)}{p(d \mid M_{\Lambda})} \right),
\end{equation}
where $M_{\Lambda}$ represents the $\Lambda$CDM model and $M_i$ denotes the $R_h=ct$ model taken into consideration. Here, $p(d \mid M)$ is Bayesian evidence, which quantifies the probability of obtaining the observed data $d$ given a model $M$. The evidence is estimated using the nested sampling algorithm implemented in the \texttt{PolyChord} library, which efficiently evaluates both the evidence and the posterior distribution.

The resulting Bayes factor provides a statistical measure of how strongly the data supports one model relative to another. Following the Jeffreys' scale~\cite{jeffreys1998theory}, the strength of evidence is interpreted as follows: $\ln B_{\Lambda i}<1$ indicates inconclusive evidence, $1<\ln B_{\Lambda i}<2.5$ corresponds to weak support, $2.5<\ln B_{\Lambda i}<5$ suggests moderate support, and $\ln B_{\Lambda i}>5$ signifies strong support for the model. A negative value of $\ln B_{\Lambda i}$ implies a statistical preference for the $\Lambda$CDM model over the corresponding $R_h=ct$ model. This qualitative interpretation provides a useful benchmark for model comparison, analogous to the interpretation of information-criterion differences discussed later.

This approach naturally balances model fit and complexity, ensuring that additional parameters are favored only when they lead to a significantly improved explanation of the data. 

We also performed a $\chi^2$ analysis to further validate our results. The $\chi^2$ statistic quantifies the deviation between the theoretical predictions and the observational data. The minimum value, $\chi^2_{min}$, corresponds to the $\chi^2$ evaluated at the best-fitting parameter set obtained from the MCMC analysis, i.e. the parameter values that maximize the posterior likelihood and provide the smallest discrepancy between the model predictions and the observations. 
In addition, we compute the reduced chi-square, defined as $\chi^2_{Red} = \chi^2_{min}/D.O.F.$, where the number of degrees of freedom is given by $D.O.F. = N - p$. Here $N$ represents the total number of observational data points and $p$ denotes the number of free parameters in the model. A value of $\chi^2_{Red} \approx 1$ indicates a statistically good fit, while significantly larger values suggest a poor fit or model inadequacy. Conversely, values much smaller than unity may indicate overfitting or overestimated observational uncertainties~\cite{andrae2010and}.
\\In addition to the Bayesian evidence and $\chi^2$ statistics, we also employ two widely used information criteria for model comparison: the Akaike Information Criterion ($AIC$)~\cite{akaike1974} and the Bayesian Information Criterion ($BIC$)~\cite{bayesian1978}. These criteria provide a balance between the goodness of fit and the complexity of the model by penalizing the inclusion of additional parameters. The $AIC$ is defined as,
\begin{equation}
    AIC = \chi^2_{min}+2p,
\end{equation}
Similarly, the $BIC$ is given by 
\begin{equation}
    BIC = \chi^2_{min} + p\ln N, 
\end{equation}
For model comparison, the differences $\Delta AIC$ and $\Delta BIC$ are computed relative to the reference $\Lambda$CDM model. Smaller values of $AIC$ or $BIC$ indicate a preferred model.
The differences are defined as
$\Delta {AIC} = {AIC}_i - {AIC}_{\Lambda}$ and 
$\Delta {BIC} = {BIC}_i - {BIC}_{\Lambda}$, 
where the $\Lambda$CDM model is taken as the reference. These information-criterion differences play a role analogous to the Jeffreys’ scale used for interpreting Bayes factors discussed earlier.
In general, $\Delta \lesssim 2$ indicates that the two models provide comparable fits to the data, 
$4 \lesssim \Delta \lesssim 7$ suggests moderate evidence against the model with the larger value, and $\Delta > 10$ is commonly interpreted as strong evidence against that model~\cite{burnham2002model,kass1995bayes}. 
This criterion is applied to both $\Delta AIC$ and $\Delta BIC$ when assessing the relative performance of the $R_h = ct$ model with respect to $\Lambda$CDM.

\begin{figure}[!htbp]
    \centering
    \includegraphics[width=1.\linewidth]{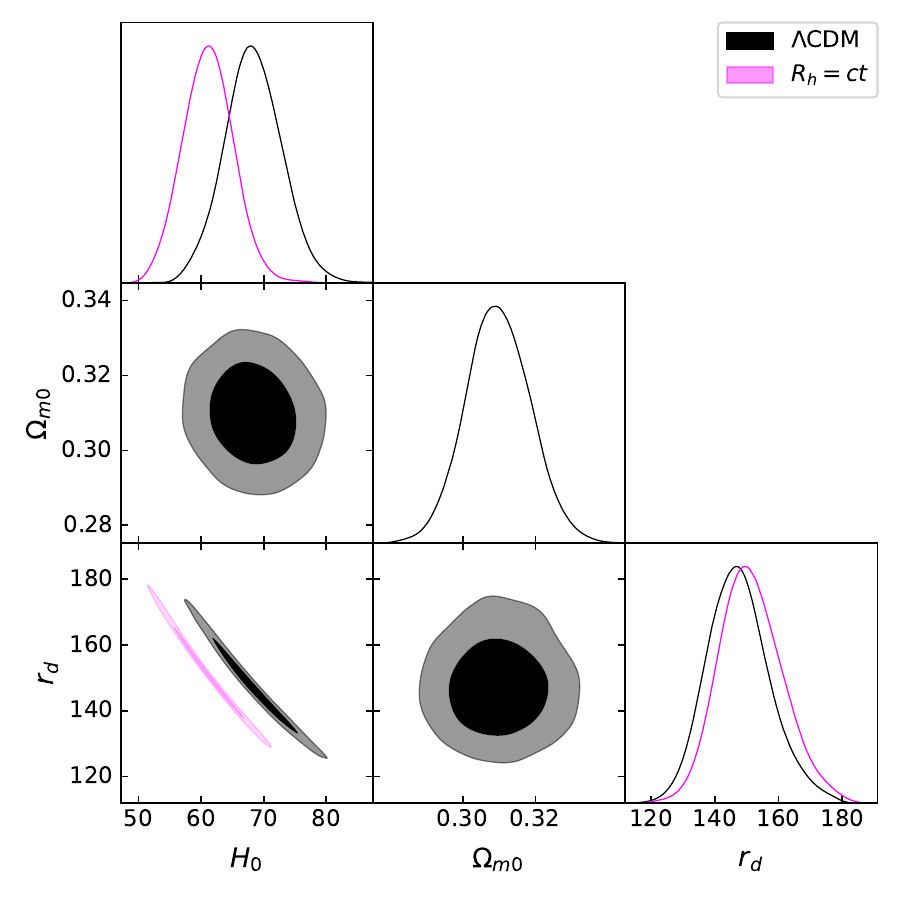}
    \caption{Combined MCMC corner plot showing parameter constraints for the $\Lambda$CDM and $R_h=ct$ cosmological models. Diagonal panels display marginalized one-dimensional posterior distributions, while off-diagonal panels show two-dimensional correlations with 68$\%$ and 95$\%$ confidence contours.}
    \label{fig_1}
\end{figure}

The following subsections summarize additional diagnostics derived from the MCMC constraints using the combined dataset (CC~+~Pantheon$^+$~+~DESI DR2), focusing on the redshift evolution of the cosmographic quantities, the cosmic age of the Universe, and the consistency of models with the DESI DR2 BAO observables.

\begin{table*}
\resizebox{\textwidth}{!}{%
\begin{tabular}{lcccccccccc}
\toprule
\textbf{Model} & $H_0$ ($\mathrm{km\,s^{-1}\,Mpc^{-1}}$) & $\Omega_{m0}$ & $r_d$ (Mpc) & $\ln{BF}$ & $\chi^2_{min}$ & $\chi^2_{red}$ & AIC & BIC & $\Delta AIC$ & $\Delta BIC$\\
\midrule
\textbf{Flat $\Lambda$CDM} & $68.3 \pm 4.6$ & $0.309 \pm 0.008$ & $147.5^{+8.6}_{-11}$ & 0 & 1574.88 & 0.975 & 1580.88 & 1597.05 & 0 & 0 \\
\addlinespace[0.2cm]
$\mathbf{R_h=ct}$ & $61.1 \pm 4.1$ & --- & $151.4^{+8.9}_{-11}$ & -2.3 & $1629.51$ & $1.008$ & $1633.51$ & $1644.29$ & 52.63 & 47.24\\
\bottomrule
\end{tabular}
}
\caption{ Posterior parameter constraints obtained from the MCMC analysis. The quoted values correspond to the marginalized posterior distributions with 68\% ($1\sigma$) credible intervals. For parameters with asymmetric posterior distributions, the uncertainties are reported using the 16$^{th}$ and 84$^{th}$ percentiles.}\label{tab_1}
\end{table*}

\subsection{Redshift evolution of $H(z)$ and $q(z)$}
A direct way to test the compatibility of $R_h=ct$ model with observations is by comparing their predictions for the Hubble parameter, $H(z)$, against the standard $\Lambda$CDM model using the most recent (CC) dataset. The expression for the Hubble parameter for the case of $\Lambda$CDM model is given as
\begin{equation}
    H_{\Lambda CDM}(z)= H_0\sqrt{\Omega_{m_0}(1+z)^3+(1-\Omega_{m_0})}.\label{LCDMeq}
\end{equation}
For the case of $R_h=ct$ model, the expression for Hubble parameter is given by equation~\ref{eq_1}.

Here, $H_0$ denotes the present day Hubble constant and $\Omega_{m_0}$ denotes the present day matter density parameter which is the fraction of the total density of the Universe contributed by matter at the current epoch ($z=0$). The estimated values of these parameters are listed in Table~\ref{tab_1} which were obtained by MCMC analysis that combines all three datasets discussed earlier in this section. Using these best-fit parameter values, we compute and plot $H_{\mathrm{model}}(z)$ for each case and directly compare the results with the observed CC data. We then proceed to analyze the DP, $q(z)$, which serves as a key cosmographic quantity to understand the expansion dynamics of the Universe. 

The DP is a dimensionless quantity that characterizes the rate of change of the cosmic expansion and help distinguish between accelerating and decelerating phases of the Universe. It is defined in terms of the scale factor $a(t)$ as, 
\begin{equation}
    q(z) = -\frac{\ddot{a}a}{\dot{a}^2} = -1 + \frac{d}{dt}\left(\frac{1}{H}\right),
\end{equation}
where $H=\dot{a}/a$ is the Hubble parameter. Expressed as a function of redshift $z$, the DP can be written as, 
\begin{equation}
    q(z) = -1 + (1+z)\frac{1}{H(z)}\frac{dH(z)}{dz},
\end{equation}
which links the expansion dynamics directly to observable quantities such as $H(z)$~\cite{visser2004jerk,cattoen2008cosmographic,capozziello2008cosmography,aviles2012cosmography,capozziello2018cosmographic,capozziello2019extended}.

A positive value of $q(z)$ signifies that the Universe is in a decelerating phase, whereas a negative value indicates accelerated expansion. The redshift at which $q(z)$ changes sign marks the transition from the matter-dominated era, where gravity slowed the expansion to the present epoch of acceleration driven by dark energy. This transition redshift, $z_{tr}$, serves as an important cosmographic marker, revealing when dark energy began to dominate the cosmic dynamics. Cosmography, in this context, provides a model-independent framework to trace the expansion history of the Universe without relying on any specific dark energy model or equation of state. 
\\For $R_h=ct$ cosmology, the scale factor evolves linearly with cosmic time $t$; $a(t)\propto t$, which implies $H= \dot{a}/a=1/t$. 
Hence, for the case of $R_h=ct$ Universe, $q(z)=0$, indicating that the expansion proceeds at a constant rate, without acceleration or deceleration.
\\It is important to note that, unlike the $\Lambda$CDM model, the $R_h=ct$ Universe predicts a constant expansion rate with $q(z)=0$ at all redshifts, and therefore does not exhibit a transition between decelerating and accelerating phases. The comparison of the deceleration parameter shown here is therefore intended to illustrate the qualitative differences in the expansion histories predicted by the two models. In particular, while $\Lambda$CDM predicts a transition from a matter-dominated decelerating phase to a late-time accelerated expansion, the $R_h=ct$ model assumes a strictly linear expansion history. The observational constraints derived from the joint dataset allow us to assess which of these expansion histories provides a better description of the data.

\begin{figure}
\begin{subfigure}{.48\textwidth}
\includegraphics[width=\linewidth]{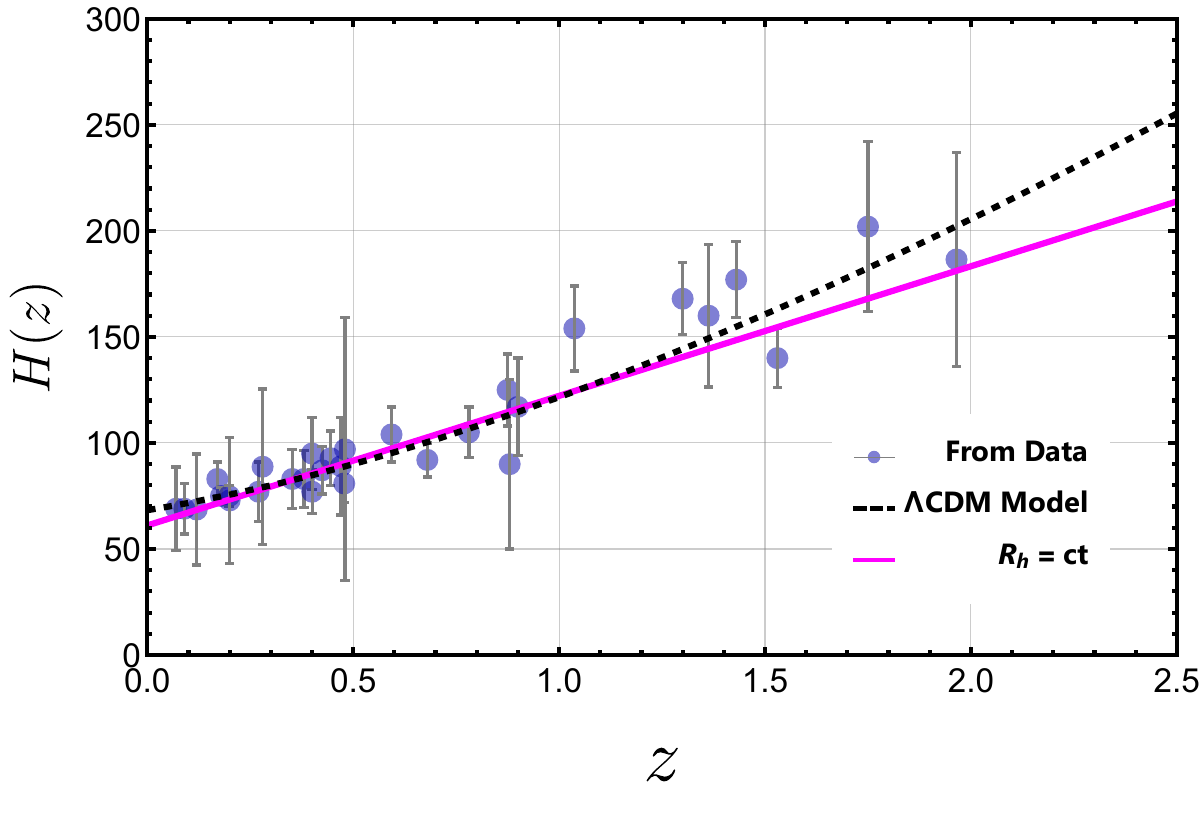}
\end{subfigure}
\hfil
\begin{subfigure}{.48\textwidth}
\includegraphics[width=\linewidth]{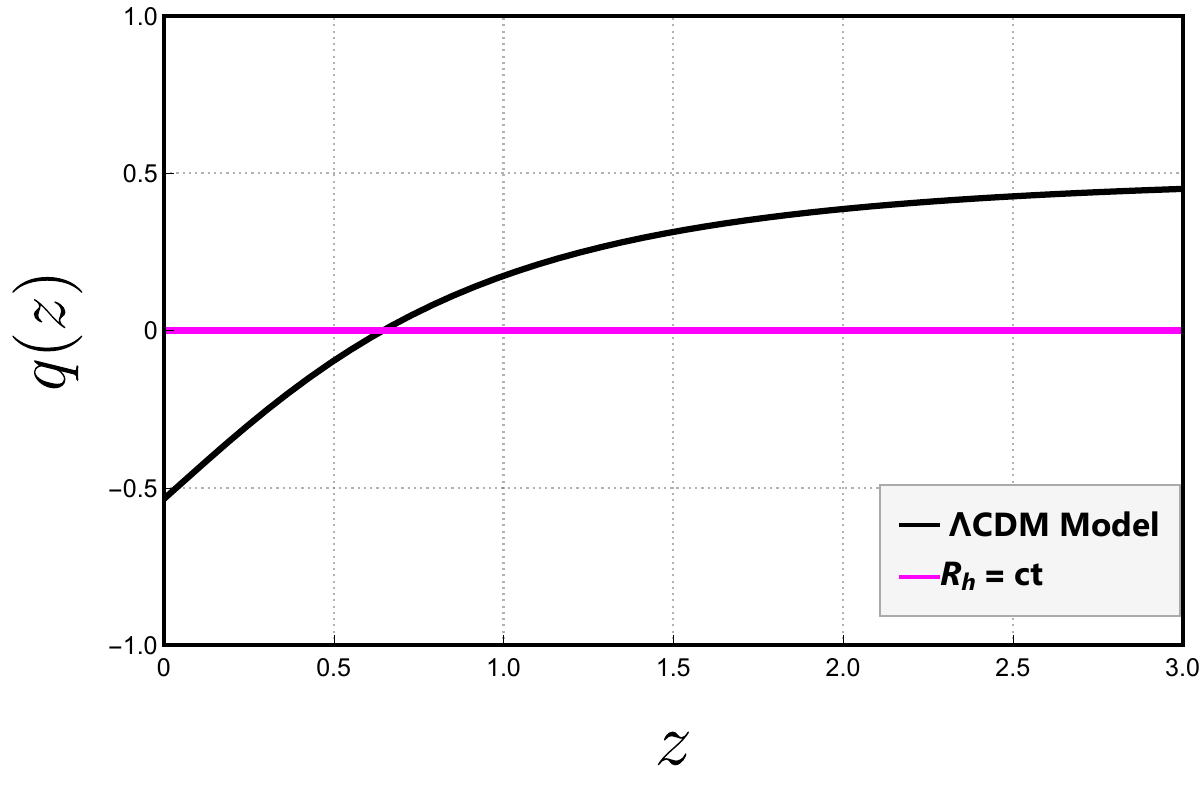}
\end{subfigure}
\caption{The top panel presents the redshift evolution of Hubble parameter ($H(z)$) with observational data comparing $R_h=ct$ model to that of $\Lambda$CDM model. The bottom panel shows the corresponding DP ($q(z)$).}\label{fig_2}
\end{figure}

\subsection{Cosmic age}
The age of the Universe constitutes a fundamental consistency test for any viable cosmological model. In a Friedmann–Lemaître–Robertson–Walker (FLRW) spacetime, the relation between cosmic time and redshift follows from $dt=-\frac{dz}{(1+z)H(z)}$, which leads to the general age-redshift relation 
\begin{equation}
t(z) = \int_{z}^{\infty} \frac{dz'}{(1+z')H(z')}.\label{eq:age_general}
\end{equation}
This expression is standard in cosmology and may be found in foundational treatments of FLRW dynamics, e.g., Peebles~\cite{peebles2020principles} and Hogg~\cite{hogg1999distance}.

For a spatially flat $\Lambda$CDM model, the Hubble function is given by eqn~\eqref{LCDMeq}
which, upon substitution into eqn~\eqref{eq:age_general}, yields the $\Lambda$CDM age–redshift relation
\begin{equation}
t(z) = \int_{z}^{\infty}
\frac{dz'}{(1+z')H(z';H_0,\Omega_m)}.
\label{eq:LCDM_age}
\end{equation}

In the present analysis, the age is not computed from fixed best-fit parameters. Instead, we propagate the full posterior distributions obtained through the MCMC analysis of the aforementioned datasets. Each MCMC sample $i$ corresponds to a distinct parameter set $(H_0^{(i)},\Omega_m^{(i)})$ and therefore produces its own realization of the age curve,
\begin{equation}
t^{(i)}(z) =
\int_{z}^{\infty}
\frac{dz'}{(1+z')H(z';H_0^{(i)},\Omega_m^{(i)})}.
\end{equation}

This procedure ensures that parameter uncertainties and correlations are fully accounted for, and it reflects the standard methodology used in modern cosmological analyses such as Planck~\cite{collaboration2020planck}, DES~\cite{abbott2022dark}, and eBOSS~\cite{alam2021completed}, where derived quantities are evaluated directly from posterior samples.

In contrast, the $R_h = ct$ cosmology is defined by the linear expansion relation, arising from the global equation-of-state constraint $\rho + 3p = 0$. In this framework, the Hubble function takes the simple form
\begin{equation}
H(z;H_0^{(i)}) = H_0^{(i)}(1+z),
\label{eq:Rhct_H}
\end{equation}
which yields the analytical age–redshift relation
\begin{equation}
t^{(i)}(z) = \frac{1}{H_0^{(i)}(1+z)}.
\label{eq:Rhct_age}
\end{equation}
The resulting temporal evolution is therefore fundamentally different from that of $\Lambda$CDM, as discussed in the original formulation of the model by Melia $\&$ Shevchuk~\cite{Melia2012}.

\begin{figure}[H]
    \centering
    \includegraphics[width=1.\linewidth]{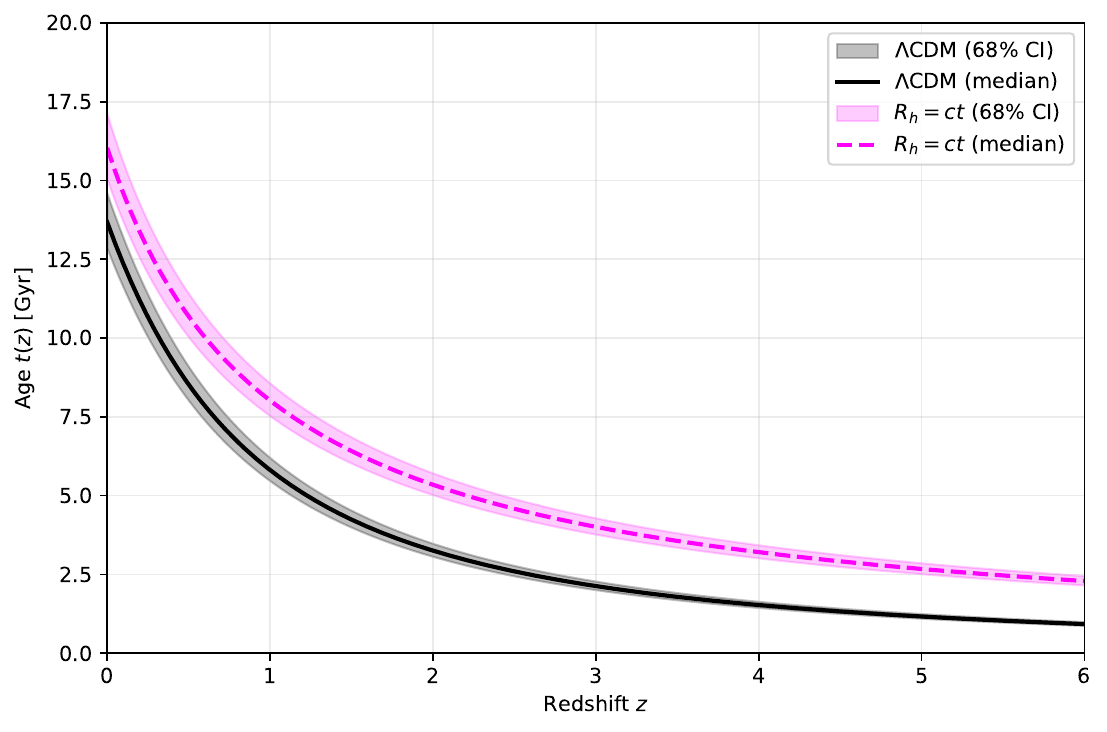}
    \caption{Age-Redshift comparison drawn between $\Lambda$CDM model and $R_h=ct$ model with 68\% confidence level. }
    \label{fig_3}
\end{figure}

\subsection{Consistency with DESI DR2 BAO observables}
To further examine the consistency of the cosmological models with large-scale structure measurements, we directly compare the BAO observables predicted by the models with the DESI DR2 data. The BAO distance ratios $D_H(z)/r_d$ and $D_M(z)/r_d$, defined by equations~\eqref{Dh_eq},~\eqref{Dm_eq} receptively, encode the transverse and radial clustering information of the galaxy distribution and provide robust geometric probes of the expansion history.
\\Using the best-fit parameters obtained from the joint MCMC analysis of the CC, Pantheon$^+$, and DESI DR2 datasets, we compute the theoretical predictions for $D_M(z)/r_d$ and $D_H(z)/r_d$ for both the $\Lambda$CDM and $R_h=ct$ cosmological models. These predictions are then compared with the corresponding DESI DR2 measurements across the redshift range covered by the survey.
\\This comparison provides a direct test of how well the models reproduce the BAO observables measured by DESI. It therefore complements the likelihood-based parameter constraints discussed earlier and offers an additional consistency check of the cosmological models against large-scale structure observations.

\begin{figure}
\begin{subfigure}{.48\textwidth}
\includegraphics[width=\linewidth]{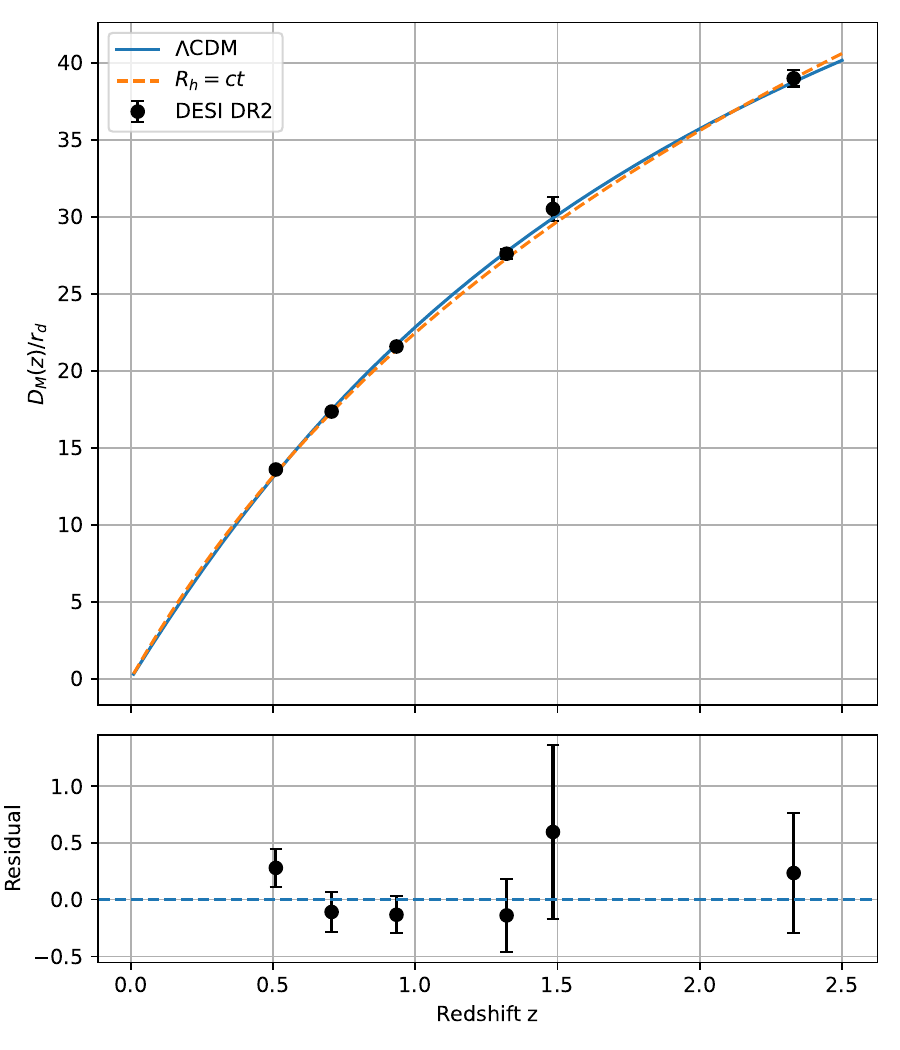}
\end{subfigure}
\hfil
\begin{subfigure}{.48\textwidth}
\includegraphics[width=\linewidth]{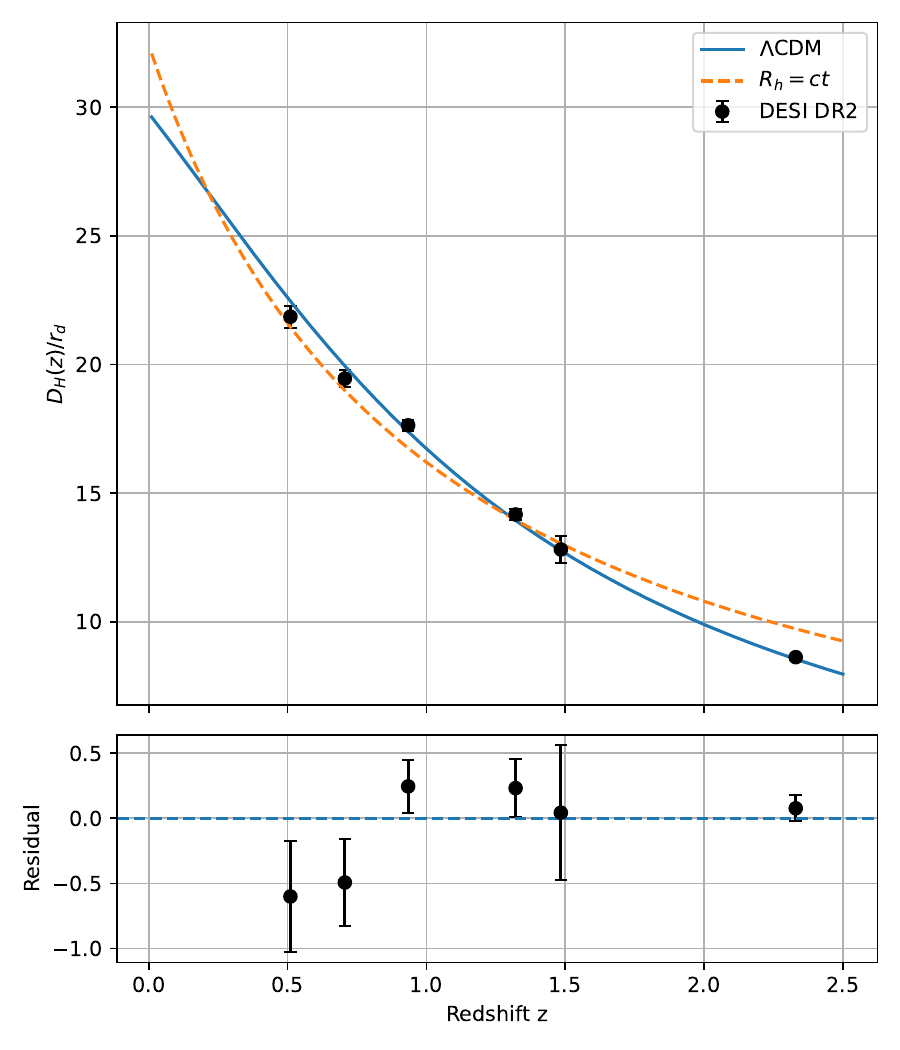}
\end{subfigure}
\caption{The upper panel shows the redshift evolution of the comoving angular diameter distance ratio $D_M(z)/r_d$, while the lower panel shows the Hubble distance ratio $D_H(z)/r_d$ for both the models under study. The data points correspond to the DESI DR2 measurements. The residual panels display the differences between the observed values and the corresponding model predictions.}\label{fig_4}
\end{figure}

The following section presents the main findings of our analysis. 

\section{Results}\label{sec_4}
Fig.~\ref{fig_1} shows the MCMC results for both $\Lambda$CDM model as well as $R_h=ct$ model in the form of corner plots. This allows one to get a direct visual comparison of the posterior distribution and parameter correlations between the two models. The diagonal panels show the marginalized one-dimensional (1D) posterior distribution for each individual parameter. The off-diagonal panels show the two-dimensional (2D) joint distribution between pairs of parameters. The contours represent the 68\% which corresponds to $1\sigma$ and 95\% which corresponds to $2\sigma$ confidence levels.

The mean values with credible intervals of 68\% $(1\sigma)$ are represented in table~\ref{tab_1} for both models. The predicted values of $H_0$ for $\Lambda$CDM model is consistent with the CC measurement reported by Moresco et al., while for the case of $R_h=ct$ model the value is lower than both Planck $(\approx67.4)${\color{red}~\cite{collaboration2020planck}} and local SH0ES $(\approx73.0)${\color{red}~\cite{Riess2022}} measurements. The values of $r_d$ align well with those obtained from Planck estimations. The associated uncertainties in the mean values of $H_0$ and $r_d$ are relatively large. The elongated (“banana-shaped”) posterior distribution visible in Fig.~\ref{fig_1} reflects the well-known degeneracy between the Hubble constant $H_0$ and the sound horizon scale $r_d$ in BAO-based analyses when $r_d$ is treated as a free parameter. In such cases, BAO measurements primarily constrain the combination $H_0 r_d$, rather than the individual parameters separately. Consequently, an increase in $H_0$ can be compensated by a corresponding decrease in $r_d$, producing the characteristic curved degeneracy observed in the parameter contours. As discussed earlier in Section~\ref{sec_3}, this degeneracy arises naturally when $r_d$ is marginalized over to maintain model independence. The inclusion of CC data in the joint analysis helps partially break this degeneracy by providing direct measurements of the expansion rate $H(z)$. The relatively large uncertainties arise from the rigorous inclusion of systematic effects in the CC dataset, where the full covariance matrix is taken into account. These uncertainties stem from several sources, including the estimation of stellar metallicities—--which can be influenced by residual young stellar populations--—variations in star formation histories, assumptions about the initial mass function (IMF), and the specific stellar libraries and population synthesis models adopted in the analysis~\cite{moresco2018setting,moresco2012new,jimenez2002constraining}. 
In contrast, the estimated values of $\Omega_{m_0}$ and $\Omega_{\Lambda_0}$ obtained for the $\Lambda$CDM model show excellent consistency with the results reported by the Planck Collaboration, which found $\Omega_{m_0} = 0.309 \pm 0.008$ and $\Omega_{\Lambda_0} = 0.691 \pm 0.008$.

The Bayesian evidence in table~\ref{tab_1}  tell the real story regarding both the models. The value of $\ln(BF)=-2.3$ for the case of $R_h=ct$ model tells that the stated model is less favored as compared to the $\Lambda$CDM model for data fitting as suggested by Jeffreys' scale. For the $R_h=ct$ model, we obtain a minimum chi-squared value of $\chi^2_{min}=1629.51$ and a reduced chi-square as $\chi^2_{Red}=1.008$. These values indicate that the model provides a reasonable fit to the data. In comparison, the $\Lambda$CDM model achieves a noticeably lower minimum chi-square value, $\chi^2_{min}=1574.88$, with $\chi^2_{Red}=0.975$, implying a stronger agreement with the observations.
The model comparison metrics reinforce this conclusion: the $\Lambda$CDM model yields $AIC=1580.88$ and $BIC=1597.05$, whereas the $R_h=ct$ model gives $AIC=1633.51$ and $BIC=1644.29$.
The resulting differences, $\Delta AIC=52.63$ and $\Delta BIC=47.24$, correspond to decisive evidence in favor of the $\Lambda$CDM model under standard model-selection criteria $(\Delta >10)$. Overall, while the $R_h=ct$ model can reproduce the data within acceptable statistical limits, the $\Lambda$CDM model offers a substantially better fit and remains the preferred cosmological description according to all evaluated criteria.

The top panel of Fig.~\ref{fig_2} displays the redshift evolution of the Hubble parameter $H(z)$ for both the $\Lambda$CDM and $R_h=ct$ cosmological models, together with the observational data from CC dataset. The $\Lambda$CDM curve reproduces the observed trend of $H(z)$ remarkably well across the entire redshift range, capturing both the low and high redshift behavior of the expansion rate. In contrast, the $R_h=ct$ model systematically predicts lower values of $H(z)$, especially at intermediate and higher redshifts $(z\gtrsim1)$, indicating that it underestimates the expansion rate of the Universe in these regimes. This deviation is consistent with the lower best-fit value of $H_0$ obtained for the $R_h=ct$ model $(61.1\pm4.1~kms^{-1}Mpc^{-1})$ compared to the $\Lambda$CDM prediction. Overall, the comparison further confirms that the $\Lambda$CDM model provides a more accurate description of the observed expansion history, in agreement with the results inferred from the chi-square and information-criterion analyses. 
At redshifts $z \gtrsim 1$, the $R_h=ct$ model shows a noticeable deviation from several of the CC measurements. However, it is important to note that the CC data at these redshifts are relatively sparse and are associated with comparatively larger uncertainties. Consequently, while the visual comparison suggests a growing discrepancy at higher redshift, the statistical significance of this deviation remains limited. 
The quantitative assessment of the model performance is therefore provided by the joint likelihood analysis using the combined CC, Pantheon$^+$, and DESI DR2 datasets, as summarized in Table~\ref{tab_1}. These results indicate that although the $R_h=ct$ model can reproduce the observational data within acceptable statistical limits, the $\Lambda$CDM model provides a better overall fit to the current data.
\\The bottom panel of Fig.~\ref{fig_2} displays the redshift evolution of the deceleration parameter $q(z)$ for the $\Lambda$CDM and $R_h=ct$ cosmological models. In the case of the $\Lambda$CDM model, the Universe undergoes a clear transition from a decelerating phase at early times $(q>0)$ to an accelerating phase at lower redshifts $(q<0)$, with the transition occurring at $z\approx 0.6$. This behavior is consistent with the standard cosmological picture in which the expansion evolves from a matter-dominated phase to a late-time dark-energy–dominated accelerated phase, as supported by multiple observational probes such as Type Ia supernovae, BAO measurements, and CMB observations~\cite{collaboration2020planck}.
In contrast, the $R_h=ct$ framework assumes that the total equation of state of the cosmic fluid satisfies $\omega=-1/3$ at all times, as discussed in the Introduction, leading to a strictly linear expansion history $a(t)\propto t$~\cite{melia2012cosmic}. This condition enforces a coasting cosmology in which the expansion rate remains constant in time. The $R_h=ct$ model therefore yields a constant value of $q=0$ throughout cosmic history, implying no transition between acceleration and deceleration.
This qualitative difference highlights a key limitation of the model: it cannot reproduce the observed late-time acceleration indicated by supernovae and BAO data. The contrasting behaviors in the figure therefore reinforce the statistical results, confirming that the $\Lambda$CDM model provides a more realistic description of the Universe’s dynamical evolution. 

It should be noted that the transition redshift discussed here is not directly inferred from an independent observational measurement, but rather arises from the expansion history implied by the best-fit cosmological parameters. The observational datasets used in this work constrain the redshift evolution of the Hubble parameter and cosmological distances. From these constraints, the corresponding deceleration parameter $q(z)$ can be reconstructed for each model. 

The comparison presented here is therefore intended as a diagnostic illustration of how the two cosmological scenarios differ in their predicted expansion histories. While $\Lambda$CDM naturally produces a transition from a decelerating to an accelerating phase, the $R_h=ct$ model predicts a constant expansion rate with $q(z)=0$ at all redshifts. The observational data then determine which of these expansion histories provides the better description of the measurements.

Fig.~\ref{fig_3} displays the median age-redshift history for both cosmological models, accompanied by their $68\%$ confidence regions constructed directly from the posterior ensemble. The $\Lambda$CDM confidence band is relatively narrow, reflecting its more precisely constrained parameters, whereas the $R_h=ct$ model shows a wider uncertainty envelope at all redshifts, primarily due to the larger posterior spread in $H_0^{R_h=ct}$. For the case of $\Lambda$CDM model, the resulting posterior distribution of the present cosmic age yields $t_0^{\Lambda CDM} = 13.676_{-0.81}^{+0.92}~Gyr$ consistent with CMB-based estimates. The $R_h=ct$ model, owing to its strictly linear expansion and its lower best-fit value for $H_0$, predicts a significantly older Universe with $t_0^{R_h=ct} = 16.035_{-0.98}^{+1.09}~Gyr$.

Across the entire redshift range probed, the $R_h=ct$ model predicts substantially older ages than $\Lambda$CDM, with the divergence increasing towards higher redshifts. This behavior, visible both in the median curves and their confidence intervals, highlights the markedly different temporal evolution inherent to linear-coasting expansion. 

To further illustrate the role of DESI DR2 in the joint analysis, we compare the theoretical predictions of both cosmological models with the BAO distance measurements provided by DESI. Fig.~\ref{fig_4} shows the evolution of the dimensionless BAO distance indicators $D_M(z)/r_d$ (top panel) and $D_H(z)/r_d$ (bottom panel) together with the DESI DR2 observations. The residuals displayed at the bottom of each panel are with respect to the observational measurements. This direct visual comparison allows one to assess how well each model reproduces the BAO observables across the probed redshift range.
These quantities probe the transverse and radial components of the BAO scale and therefore provide direct geometric constraints on the expansion history of the Universe. In contrast to CC, which measure the expansion rate $H(z)$ directly, BAO observations constrain combinations of the comoving distance and the sound horizon scale. As a result, DESI measurements provide complementary information that helps tighten the joint constraints when combined with CC and Pantheon$^+$ data.
The figure shows that both models are able to reproduce the DESI measurements within the observational uncertainties, although the $\Lambda$CDM model provides a slightly better overall agreement across the full redshift range. However, a closer inspection of the residual panels reveals a qualitative difference between the two models. In particular, the $\Lambda$CDM model exhibits residuals that remain relatively small and randomly distributed around zero, indicating a consistent tracking of the data. In contrast, the $R_h=ct$ model shows a more systematic pattern, especially in the $D_H(z)/r_d$ panel, where the curve initially lies above the data, dips below it at intermediate redshifts, and rises again at higher redshifts. This structured deviation suggests that, while the model remains statistically consistent within uncertainties, it does not follow the observed trend as closely as $\Lambda$CDM across the full redshift range. The inclusion of DESI DR2 therefore plays an important role in the joint likelihood analysis by providing independent geometric constraints on the expansion history, particularly at intermediate and high redshifts.

\section{Conclusion}\label{sec_5}
In this work, we carried out a detailed statistical comparison between the linearly coasting $R_h=ct$ cosmology and the standard $\Lambda$CDM model using recent observational data from CC, DESI DR2, and Pantheon$^+$. The analysis combines MCMC parameter estimation with multiple model-selection diagnostics, including minimum $\chi^2$, information criteria, and Bayesian evidence. Through a full likelihood analysis incorporating minimum chi-square statistics, information criteria, and Bayesian evidence, we find that although both models provide acceptable fits to the data, the $\Lambda$CDM framework consistently out performs $R_h=ct$ across all statistical indicators. This result highlights the strong empirical support for the standard model in describing the expansion history of the Universe. 

One of the principal shortcomings of the $R_h=ct$ model is its inability to generate the observed transition from early-time deceleration to late-time acceleration. This limitation has been widely discussed in previous studies of the model and is confirmed by the present analysis using the combined CC, Pantheon$^+$, and DESI DR2 datasets. In our results, this behavior is reflected in the best-fit value of the Hubble constant $(H_0 = 61.1 \pm 4.1~\mathrm{km\,s^{-1}\,Mpc^{-1}})$ obtained for the $R_h=ct$ model, as well as in the mismatch between its predicted $H(z)$ evolution and the observational data at intermediate and high redshifts. In contrast, $\Lambda$CDM naturally accommodates these features through the presence of a cosmological constant, enabling it to reproduce both the magnitude and redshift evolution of the expansion rate with significantly higher fidelity. 
The cosmic age analysis further reinforces this conclusion. The posterior-derived present age of the Universe in $\Lambda$CDM is broadly consistent with the value inferred from the Planck 2018 CMB analysis~\cite{collaboration2020planck}. Although this age differs from CMB-based estimates, recent JWST observations indicating the existence of massive and evolved galaxies at redshifts $z\gtrsim10-13$~\cite{labbe2023population} have reignited discussions regarding the timeline of early structure formation. Additional analyses of old galaxy populations~\cite{conselice2025epochs} and dynamical age estimates of early-type systems~\cite{cappellari2016structure} have similarly motivated further investigation into the detailed chronology of galaxy assembly in the early Universe. While these developments are still under active investigation, they highlight the importance of continuing to test cosmological models against increasingly precise observational constraints.

Moreover, the DESI DR2 analysis~\cite{karim2025desi, Scherer:2025esj} reports that dynamical dark energy models provide a statistically superior fit to the BAO data compared to the cosmological constant, challenging one of the foundational assumptions of $\Lambda$CDM. Together with the persistent Hubble tension and the unknown physical nature of dark matter and dark energy, these results suggests that the standard model---despite its success---may not represent a complete description of cosmic evolution. 

The novelty of this work lies in presenting the first joint CC + Pantheon$^+$ + DESI DR2 statistical comparison of the $R_h=ct$ and $\Lambda$CDM cosmologies within a unified Bayesian framework. The inclusion of the recent DESI DR2 BAO measurements extends the analysis beyond earlier studies based primarily on CC or supernova datasets by incorporating precise large-scale structure distance measurements across a wide redshift range. In particular, the DESI DR2 data provide strong constraints on the redshift evolution of the BAO distance indicators, which further restrict the allowed expansion histories of the models and highlight the tension between the linear expansion predicted by $R_h=ct$ and the observed late-time acceleration. 

Our analysis also includes a full posterior-propagated cosmic-age calculation and a comprehensive model comparison based on $\chi^2$, AIC, BIC, and Bayesian evidence. Together, these elements provide a transparent assessment of the viability of the $R_h=ct$ model relative to the standard cosmological paradigm using the latest available late-time observational constraints.

\section*{Acknowledgments}

This work by A. F. S. is partially supported by National Council for Scientific and Technological
Development - CNPq project No. 312406/2023-1.

\section*{Declaration}
No new data are used here. The datasets used here are publicly available.

\bibliographystyle{apsrev4-2} 
\bibliography{references}
\end{document}